\documentclass[
floatfix,
twocolumn,
aps,
superscriptaddress,
pra,
10pt,
nofootinbib,
longbibliography
]{revtex4-2}

\usepackage[T1]{fontenc}
\usepackage[utf8]{inputenc}
\usepackage[english]{babel}
\usepackage{ulem}
\usepackage{amsmath}
\usepackage{amssymb}
\usepackage{amsfonts}
\usepackage{mathtools}
\usepackage{braket}
\usepackage{titlesec}
\usepackage[colorlinks=true,linktocpage=true]{hyperref}
\hypersetup{
    allcolors={teal!70!black}
}
\usepackage{float}

\renewcommand{\thesection}{\Roman{section}}
\renewcommand{\thesubsection}{\Alph{subsection}}
\titleformat{\section}{\normalfont\small\bfseries\centering}{\thesection.}{0.6em}{\MakeUppercase}
\titleformat{\subsection}{\normalfont\small\bfseries\centering}{\thesubsection.}{0.6em}{}
\newcommand{\FT}{\mathrm{FT}}
\newcommand{\Mk}{\mathrm{M}}
\newcommand{\ex}{\mathrm{ex}}
\newcommand{\sinc}{\operatorname{sinc}}
\newcommand{\dd}{\mathrm{d}}
\newcommand{\RQ}{\mathcal{R}_Q}
\newcommand{\Rchi}{\mathcal{R}_\chi}
\newcommand{\RK}{\mathcal{R}_K}
\newcommand{\F}{\mathcal{F}}
\newcommand{\Kt}{\mathcal{K}}
\usepackage{graphicx}
\graphicspath{{pdf/}}
\usepackage[dvipsnames]{xcolor}

\begin{document}

\title{Finite-time spectral enhancement of transient quantum thermometry}

\author{M. Tahir Naseem}
\email{mnaseem16@ku.edu.tr}
\affiliation{
Faculty of Basic Sciences,
Ghulam Ishaq Khan Institute of Engineering Sciences and Technology,
Topi 23640, District Swabi, Khyber Pakhtunkhwa, Pakistan
}
\affiliation{
National Center for Quantum Computing,
University of Engineering and Technology Narowal,
10 km Muridke Road, Near Adda Siraj, Narowal, Pakistan
}

\author{Asghar Ullah}
\email{aullah21@ku.edu.tr}
\affiliation{
Department of Physics,
Ko\c{c} University,
34450 Sar\i yer, Istanbul, T\"urkiye
}
\affiliation{
Department of Electrical \& Electronics Engineering,
Ko\c{c} University,
34450 Sar\i yer, Istanbul, T\"urkiye
}
\author{\"Ozg\"ur E. M\"ustecapl\i o\u{g}lu}
\email{omustecap@ku.edu.tr}
\affiliation{
Department of Physics,
Ko\c{c} University,
34450 Sar\i yer, Istanbul, T\"urkiye
}
\affiliation{
T\"UB\.{I}TAK Research Institute for Fundamental Sciences (TBAE),
41470 Gebze, T\"urkiye 
}
\author{Mehmet Cengiz  Onba\c{s}l\i}
\email{monbasli@ku.edu.tr}
\affiliation{
Department of Physics,
Ko\c{c} University,
34450 Sar\i yer, Istanbul, T\"urkiye
}
\affiliation{
Department of Electrical \& Electronics Engineering,
Ko\c{c} University,
34450 Sar\i yer, Istanbul, T\"urkiye
}

\date{\today}
\begin{abstract}
Transient quantum thermometry extracts temperature information before a probe reaches thermal equilibrium. We study how finite interaction times affect thermometric precision for a harmonic-oscillator probe prepared in a number state and weakly coupled to a thermal bosonic reservoir. A second-order time-convolutionless master equation retains the finite-time reservoir response, allowing modes away from the probe frequency to shape the populations and their temperature dependence. For the baseline parameters, this response raises the quantum Fisher information about fivefold above the Markovian prediction, in agreement with an exact solution of the model. When the probability of leaving the initial state is small, we derive an analytical condition for this enhancement. To assess its operational significance, we examine the temperature information acquired per unit time. Under the exact dynamics, this rate is largest within the reservoir memory time. With negligible preparation and readout overhead, the optimum is about five times the best Markovian rate and 30 times an upper bound on the equilibrium rate. The advantage survives smooth switching, although overhead and longer coupling ramps reduce it.
\end{abstract}

\maketitle

\section{Introduction}

Quantum thermometry infers temperature from measurable changes in quantum probes, offering access to microscopic
systems where conventional thermometers are difficult to implement~\cite{Stace2010,PhysRevLett.114.220405,
Brunelli2011,Brunelli2012}. Experiments with diamond spins and optomechanical systems demonstrate this approach~\cite{David,Kucsko2013,
Neumann2013,Purdy}. The quantum Fisher information bounds local estimation precision by quantifying the distinguishability
of probe states at nearby temperatures~\cite{Helstrom1969, BraunsteinCaves1994}. At equilibrium, sensitivity depends on energy fluctuations,
spectral gaps, correlations, and the accessible part of the system~\cite{Mehboudi_2015,Paris_2016, DePasquale2016,Miller2018,Campbell_2018,Hovhannisyan2018}.
Measurement restrictions~\cite{Potts2019,Jorgensen2020, PRXQuantum.2.020322,Cenni2022} and backaction~\cite{Seveso2018} further constrain
the information extracted. Global and Bayesian approaches address estimation over broader temperature ranges~\cite{Mok2021,Rubio2021},
including adaptive strategies that use prior information~\cite{Mehboudi2022Bayesian,Glatthard2022Bayesian}.
For an interacting probe, a further practical question is how much temperature information can be acquired before equilibrium is reached.

In this transient regime, thermometric performance depends on the probe preparation and interrogation time,
as explored in qubit models and photonic simulations~\cite{Jevtic2015,Mancino2017, PhysRevA.98.050101,Feyles2019}.
Ancilla-assisted schemes can encode temperature in coherences~\cite{PhysRevA.98.042124, PhysRevResearch.5.043184,PhysRevA.110.032605},
while initial and bath-induced correlations can improve sensitivity~\cite{bb24-kfrz,PhysRevLett.128.040502}.
Collisional protocols~\cite{PhysRevLett.123.180602,Alves2024} and sequential measurements~\cite{PhysRevApplied.22.024069} also exploit information acquired through successive
interactions or observations. The time required to acquire information is addressed by finite-time precision bounds~\cite{Sekatski2022},
prethermal probes~\cite{AntoSztrikacs2024}, and nonequilibrium single-atom sensing~\cite{Bouton2020}. For oscillator probes, nonequilibrium operation
does not necessarily improve sensitivity~\cite{PhysRevResearch.2.033498}, although suitable nonclassical Gaussian
preparations offer advantages~\cite{Mirkhalaf2024}. Fock states and other non-Gaussian preparations can further improve transient performance under dissipative
dynamics~\cite{UllahNaseemMustecaplioglu2025,gsh7-r7ms}. Exact treatments of Gaussian
oscillator probes have investigated strong-coupling finite-time thermometry~\cite{winczewski2026nonequilibriumquantumthermometrybosonic}. Alongside these preparation-dependent advantages, the reservoir response determines how temperature
information is encoded during the interaction.

The reservoir influences thermometric performance through the frequencies and coupling strengths of
its modes. Impurity probes access temperature information in Bose and Fermi gases~\cite{Hangleiter2015,
Mehboudi2019PolaronThermometry,Johnson2016, PhysRevLett.125.080402,PhysRevA.109.023309}, while dephasing schemes connect sensitivity to
the reservoir spectrum~\cite{Razavian2019,PhysRevA.101.032112}. Strong coupling can improve thermometry, but its
benefits depend on the operating regime~\cite{PhysRevA.96.062103,Ravell2024,t74h-c8kw}. Reservoir effects have also been investigated
in non-Markovian models~\cite{PhysRevApplied.17.034073, PhysRevE.110.024132,Aiache2025Structured,c87r-22mn},
and periodic driving can control the frequencies involved in sensing~\cite{MukherjeeEtAl2019Thermometry,
Glatthard2022bendingrulesoflow}. These studies motivate examining how reservoir modes contribute to temperature sensing during a finite probe–reservoir interaction.
For the oscillator model considered here, the Markovian rates sample the reservoir only at the probe frequency. During a finite interaction, however, modes away from resonance can also influence the temperature dependence of the probe populations. The non-Markovian studies cited above show that reservoir memory can change thermometric precision. Our aim is different. We identify which reservoir frequencies carry the extra information, give a closed condition for when it raises the quantum Fisher information (QFI), and test whether the gain survives as an operational advantage.

We address these questions with a harmonic-oscillator probe prepared in a number (Fock) state and weakly coupled to a bosonic reservoir. We describe its dynamics with a second-order time-local master equation that retains the finite interaction interval in the reservoir response. We also solve the model exactly and use the exact solution as a benchmark. Modes away from the probe frequency affect both the excitation probabilities and how strongly they change with temperature. These modes can raise the QFI above the Markovian prediction at the same interrogation time. When the probability of leaving the initial state is small, an analytical QFI ratio identifies when this happens.
Because the finite-time and Markovian QFIs describe the same experiment, their ratio measures what the Markovian description misses. We therefore ask a second, operational question. We compare the information acquired per unit interrogation time under the exact dynamics with the best rate available after the reservoir memory has decayed. We find that the optimal interrogation lies inside the memory time. We then show how preparation overhead and a finite switching time limit this advantage, and we map its dependence on temperature, initial number state, and reservoir spectrum.

The remainder of this work is organized as follows. Section~\ref{sec:model} introduces the model, its finite-time and Markovian dynamics, and its exact solution. Section~\ref{sec:results} defines the QFI comparison, develops the enhancement mechanism, and tests it numerically. It then examines the information rate, finite switching times, and the dependence on temperature and initial state. Section~\ref{sec:conclusion} concludes the work. Appendices~\ref{app:me}--\ref{app:switch} give the master-equation derivation, the analytical population and QFI calculations, the numerical validation and spectral scans, the exact solution, and the switching filte

\section{Model and finite-time reservoir response}\label{sec:model}

We consider a harmonic-oscillator probe coupled to a thermal bosonic reservoir for a controlled interaction time. The reservoir temperature is inferred from the probe's number-state populations. We introduce the physical model and its finite-time dynamics, together with the Markovian benchmark used for comparison.

\subsection{Physical model}\label{sec:physmodel}

The probe, reservoir, and their interaction are described by
\begin{align}
H &= H_S + H_B + H_I, \label{eq:H}\\
H_S &= \hbar\omega_0 a^\dagger a, \\
H_B &= \sum_k \hbar\omega_k b_k^\dagger b_k, \\
H_I &= \hbar\lambda\sum_k\left(g_k a^\dagger b_k + g_k^* a b_k^\dagger\right). \label{eq:HI}
\end{align}
Here, $H$ is the total Hamiltonian, while $H_S$, $H_B$, and $H_I$ describe the probe, reservoir, and interaction, respectively. The operators $a$ and $b_k$ annihilate excitations of the probe and reservoir mode $k$, respectively, and satisfy $[a,a^\dagger]=1$ and $[b_k,b_\ell^\dagger]=\delta_{k\ell}$, where $\delta_{k\ell}$ is the Kronecker delta. The probe and reservoir modes have positive angular frequencies $\omega_0$ and $\omega_k$, respectively. The coupling amplitude $g_k$ has units of angular frequency, and $\lambda\ll1$ is a dimensionless coupling parameter. Each interrogation begins from the factorized state
\begin{equation}
\rho_{\rm tot}(0)=\rho(0)\otimes\rho_B(T), \label{eq:init}
\end{equation}
\begin{equation}
\rho(0)=|n_0\rangle\langle n_0|,\qquad \rho_B(T)=\frac{e^{-H_B/(k_BT)}}{Z_B}.
\end{equation}
Here, $\rho_{\rm tot}$ is the joint state, $\rho$ is the reduced probe state, and $\rho_B(T)$ is the thermal reservoir state at temperature $T$. The Fock state $|n_0\rangle$ has a prescribed nonnegative integer occupation $n_0$. The constant $k_B$ is Boltzmann's constant, and $Z_B=\mathrm{Tr}_B[e^{-H_B/(k_BT)}]$ is the reservoir partition function. Equation~\eqref{eq:init} assumes that the probe and reservoir are initially uncorrelated.

After preparation, the probe interacts with the reservoir for a time $t$, and its excitation number is measured. Until Sec.~\ref{sec:switch} we take the coupling to switch on at time zero and off at time $t$ instantaneously. Section~\ref{sec:switch} replaces this step with a smooth switching profile. Independent repetitions under the same initial conditions sample the number-state populations used for temperature inference. The finite-time and Markovian evolutions are compared using the same initial preparation, interrogation time, and physical parameters. Throughout the analysis, temperature derivatives are taken at fixed interrogation time, initial occupation, probe frequency, and reservoir spectral and coupling parameters.

\subsection{Dynamics and Markovian benchmark}

For the Hamiltonian in Eq.~\eqref{eq:H} and the initial state in Eq.~\eqref{eq:init}, we obtain the reduced dynamics from a second-order time-convolutionless (TCL2) expansion that retains the reservoir response over the finite interaction interval~\cite{breuer2002theory}. The derivation in Appendix~\ref{app:me} gives the Schr\"odinger-picture master equation
\begin{align}
\frac{\dd\rho(T,t)}{\dd t}={}&-\frac{i}{\hbar}\left[H_S+H_{LS}(T,t),\rho(T,t)\right]\nonumber\\
&+\gamma_{-,\FT}(T,t)\mathcal{D}[a]\rho(T,t)\nonumber\\
&+\gamma_{+,\FT}(T,t)\mathcal{D}[a^\dagger]\rho(T,t). \label{eq:me}
\end{align}
Here, the Lindblad dissipator is
\begin{equation}
\mathcal{D}[L]\rho=L\rho L^\dagger-\frac12\left\{L^\dagger L,\rho\right\},
\end{equation}
where $L$ is the jump operator and braces denote an anticommutator. The reservoir-induced Lamb-shift Hamiltonian $H_{LS}(T,t)$ is given in Eq.~\eqref{eq:HLS}. The time-local coefficients $\gamma_{-,\FT}$ and $\gamma_{+,\FT}$ describe downward and upward transitions, respectively, with FT denoting the finite-time response. Their expressions in Eq.~\eqref{eq:A14} can be written in the equivalent sinc form
\begin{align}
\gamma_{-,\FT}(T,t)={}&2\lambda^2t\int_0^\infty\!\dd\omega\,J(\omega)\left[n_T(\omega)+1\right]\nonumber\\
&\times\sinc[(\omega_0-\omega)t], \label{eq:gm}\\
\gamma_{+,\FT}(T,t)={}&2\lambda^2t\int_0^\infty\!\dd\omega\,J(\omega)n_T(\omega)\nonumber\\
&\times\sinc[(\omega_0-\omega)t]. \label{eq:gp}
\end{align}
Here, $J(\omega)$ is the reservoir spectral density, $n_T(\omega)$ is the thermal occupation at angular frequency $\omega$, and $\sinc x=\sin x/x$, with $\sinc0=1$. The spectral density and thermal occupation are
\begin{align}
J(\omega)&=\sum_k|g_k|^2\delta(\omega-\omega_k), \label{eq:J}\\
n_T(\omega)&=\frac{1}{\exp[\hbar\omega/(k_BT)]-1},
\end{align}
where $\delta$ is the Dirac delta distribution. At finite $t$, the sinc kernels in Eqs.~\eqref{eq:gm} and \eqref{eq:gp} allow off-resonant reservoir modes to contribute to the transition coefficients.

To specify the reservoir coupling spectrum associated with the interaction Hamiltonian in Eq.~\eqref{eq:HI}, we choose the continuum form~\cite{Reichert2016}
\begin{equation}
J(\omega)=\kappa\frac{\Gamma\Omega^2\omega}{(\omega^2-\Omega^2)^2+(\Gamma\omega)^2}. \label{eq:Jform}
\end{equation}
Here, $\Omega$ sets the characteristic reservoir resonance frequency, $\Gamma$ controls its spectral width, and $\kappa$ scales the spectral coupling weight. This form is motivated by a harmonic resonance broadened through coupling to an Ohmic background~\cite{Reichert2016}, where $\Gamma$ describes the damping of the collective resonance. Through Eq.~\eqref{eq:J}, it specifies how the squared microscopic couplings are distributed over the reservoir normal-mode frequencies. The spectral maximum depends on both $\Omega$ and $\Gamma$, and need not coincide with $\Omega$. This choice allows the spectral position, width, and coupling scale to be varied relative to the fixed probe frequency. The reservoir correlations decay on the memory time $\tau_B\simeq2/\Gamma$, which equals about $17/\omega_0$ for the
baseline width $\Gamma/\omega_0=0.12$.


The Markovian benchmark follows from the long-time limit of the kernel $t\,\sinc[(\omega_0-\omega)t]$ appearing in Eqs.~\eqref{eq:gm} and \eqref{eq:gp}. For fixed $\omega_0$,
\begin{equation}
\frac{\sin[(\omega_0-\omega)t]}{\omega_0-\omega}\xrightarrow[t\to\infty]{\mathcal{D}}\pi\delta(\omega-\omega_0),
\end{equation}
where the arrow denotes distributional convergence under integration against suitable smooth frequency weights. Applying this limit to Eqs.~\eqref{eq:gm} and \eqref{eq:gp} gives
\begin{equation}
\gamma_{-,\Mk}(T)=\gamma_0\left[\bar n(T)+1\right],\qquad \gamma_{+,\Mk}(T)=\gamma_0\bar n(T), \label{eq:markov}
\end{equation}
where M denotes the Markovian benchmark. The damping scale and resonant thermal occupation are
\begin{equation}
\gamma_0=2\pi\lambda^2J(\omega_0),\qquad \bar n(T)=n_T(\omega_0). \label{eq:g0}
\end{equation}
Here, $\gamma_0$ is temperature independent, and $\bar n(T)$ is the reservoir occupation at the probe frequency. The Markovian rates sample the reservoir spectrum at $\omega_0$, whereas the finite-time coefficients $\gamma_{\pm,\FT}$ include off-resonant contributions.

\subsection{Exact solution}\label{sec:exact}

The model in Eqs.~\eqref{eq:H}--\eqref{eq:HI} conserves the total excitation number and is quadratic. Its Heisenberg equations are therefore linear and can be solved exactly. Appendix~\ref{app:exact} shows that the probe operator evolves as
\begin{equation}
a(t)=G(t)\,a+\sum_k f_k(t)\,b_k, \label{eq:aexact}
\end{equation}
where $G(t)$ is the probe propagator and $f_k(t)$ are reservoir amplitudes. Two functions then fix the reduced dynamics. The first is the transmissivity $\eta(t)=|G(t)|^2$. The second is the thermal excitation transferred into the probe, $N(T,t)=\sum_k|f_k(t)|^2n_T(\omega_k)$. Only $N$ depends on temperature. The exact populations obey Eq.~\eqref{eq:B1} with the exact time-local rates
\begin{equation}
\gamma_{+,\ex}=\dot N+\Gamma_\ex N,\qquad \gamma_{-,\ex}=\Gamma_\ex+\gamma_{+,\ex}, \label{eq:exrates}
\end{equation}
where $\Gamma_\ex(t)=-\dot\eta/\eta$ and dots denote time derivatives. We use this solution as a benchmark for the TCL2 results in Sec.~\ref{sec:results}.

\section{Finite-time thermometric enhancement}\label{sec:results}

We now examine how these off-resonant contributions affect the temperature information carried by the probe. We first introduce the QFI and show that excitation-number measurements attain it. We then relate the QFI to the reservoir-induced population changes and their temperature dependence. Finally, we compare the analytical prediction of QFI enhancement over the Markovian benchmark with numerical results.

\subsection{Thermometric figure of merit}

We quantify thermometric performance through the precision attainable from repeated measurements of the probe. For $m$ independent, identically prepared interrogations of duration $t$, the quantum Cram\'er--Rao bound gives~\cite{Helstrom1969,Paris2009}
\begin{equation}
\Delta^2T\geq\frac{1}{m\F_X(T,t)}, \label{eq:QCRB}
\end{equation}
where $\Delta^2T$ is the variance of a locally unbiased temperature estimator and $\F_X(T,t)$ is the QFI per interrogation for temperature estimation. The index $X\in\{\FT,\Mk\}$ distinguishes the finite-time and Markovian evolutions. At fixed $m$, a larger QFI lowers the corresponding variance bound. To evaluate the QFI from the probe state, we introduce the symmetric logarithmic derivative,
\begin{align}
\partial_T\rho_X(T,t)&=\frac12\left[\rho_X(T,t)L_{T,X}+L_{T,X}\rho_X(T,t)\right],\nonumber\\
\F_X(T,t)&=\mathrm{Tr}\left[\rho_X(T,t)L_{T,X}^2\right], \label{eq:SLD}
\end{align}
where $\rho_X(T,t)$ is the reduced probe state under evolution $X$, and $L_{T,X}$ is the Hermitian symmetric logarithmic derivative, with its dependence on $T$ and $t$ left implicit. Temperature derivatives follow the fixed-parameter convention specified in Sec.~\ref{sec:physmodel}.

The Hamiltonian terms in Eq.~\eqref{eq:me} are diagonal in the number basis, and the dissipators in Eq.~\eqref{eq:me} preserve diagonal states. The initial Fock state therefore evolves into
\begin{equation}
\rho_X(T,t)=\sum_{n=0}^\infty p_n^X(T,t)|n\rangle\langle n|, \label{eq:diag}
\end{equation}
where $p_n^X(T,t)$ is the probability of observing $n$ probe excitations. Because the number basis is independent of temperature, differentiating Eq.~\eqref{eq:diag} changes only the populations, and $L_{T,X}$ can be chosen diagonal. Substituting this state into the first line of Eq.~\eqref{eq:SLD} and taking number-state matrix elements gives $\langle n|L_{T,X}|n\rangle=\partial_Tp_n^X/p_n^X$. Inserting these matrix elements into the second line of Eq.~\eqref{eq:SLD} yields
\begin{equation}
\F_X(T,t)=\sum_{n:\,p_n^X>0}\frac{\left[\partial_Tp_n^X(T,t)\right]^2}{p_n^X(T,t)}. \label{eq:QFI}
\end{equation}
The sum runs over number states with nonzero probability. Ideal number-resolved detection samples the distribution in Eq.~\eqref{eq:diag}. Its classical Fisher information is the population sum in Eq.~\eqref{eq:QFI}, giving
\begin{equation}
\F_{{\rm num},X}(T,t)=\F_X(T,t),
\end{equation}
where $\F_{{\rm num},X}$ denotes the classical Fisher information of the number-measurement outcomes. Thus, number measurement attains the QFI in a basis independent of the unknown temperature. Under the usual regularity conditions, an asymptotically efficient estimator can approach the variance bound in Eq.~\eqref{eq:QCRB} as the number of independent repetitions increases.

Having expressed the QFI in terms of the measured populations, we now compare the finite-time dynamics with the Markovian benchmark. At the same temperature and interrogation time, we define
\begin{equation}
\RQ(T,t)=\frac{\F_\FT(T,t)}{\F_\Mk(T,t)}. \label{eq:RQ}
\end{equation}
Here, $\RQ$ quantifies the effect of the Markovian description on the QFI by comparing the finite-time and Markovian results under the same conditions. Both QFIs use the same initial Fock state, probe frequency, reservoir parameters, temperature, and interrogation time. Values $\RQ>1$ indicate that the Markovian description predicts a lower QFI, whereas $\RQ<1$ indicate that it predicts a higher QFI. Both QFIs describe the same protocol, so $\RQ$ measures the information that the Markovian description misses. 
Evaluating Eq.~\eqref{eq:RQ} requires the populations and their temperature derivatives. We next determine how transitions change these populations during the interrogation and which reservoir frequencies contribute to their temperature dependence.

\subsection{Spectral mechanism of enhancement}\label{sec:mech}

To explain how finite-time reservoir response can enhance the QFI, we first identify how the transition probabilities and their temperature dependence contribute to it. We consider an interrogation during which the probability of leaving the initial Fock state is small. Therefore leading population changes then involve only the neighbouring number states. Expanding the population dynamics obtained from Eq.~\eqref{eq:me} to leading order gives
\begin{align}
p^\FT_{n_0-1}(T,t)&\simeq n_0\Kt_{-,\FT}(T,t),\nonumber\\
p^\FT_{n_0}(T,t)&\simeq1-\varepsilon_\FT(T,t),\nonumber\\
p^\FT_{n_0+1}(T,t)&\simeq(n_0+1)\Kt_{+,\FT}(T,t). \label{eq:pop}
\end{align}
Here, the integrated transition coefficients are
\begin{equation}
\Kt_{\pm,\FT}(T,t)=\int_0^t\gamma_{\pm,\FT}(T,s)\,\dd s, \label{eq:K}
\end{equation}
and the leading probability of leaving the initial state is
\begin{equation}
\varepsilon_\FT=n_0\Kt_{-,\FT}+(n_0+1)\Kt_{+,\FT}. \label{eq:eps}
\end{equation}
For $n_0=0$, the downward-transition outcome is absent. The expansion requires $\varepsilon_\FT\ll1$, in addition to the weak-coupling approximation used to obtain the master equation~\eqref{eq:me}. Its derivation and validity are discussed in Appendix~\ref{app:pop}.

Differentiating the populations in Eq.~\eqref{eq:pop} with respect to temperature, substituting the results into Eq.~\eqref{eq:QFI}, and retaining the leading contributions gives
\begin{equation}
\F^{\rm lead}_\FT=\chi_\FT^2\left[\frac{n_0}{\Kt_{-,\FT}}+\frac{n_0+1}{\Kt_{+,\FT}}\right], \label{eq:Flead}
\end{equation}
where lead denotes the leading weak-transition expression and
\begin{equation}
\chi_\FT(T,t)=\partial_T\Kt_{-,\FT}(T,t)=\partial_T\Kt_{+,\FT}(T,t). \label{eq:chi}
\end{equation}
The response $\chi_\FT$ is the temperature derivative of the integrated transition coefficients. Equation~\eqref{eq:Flead} omits the higher-order contribution from the initially occupied state. The temperature derivatives of all three retained populations are of the same order, but the initial-state probability remains close to unity. Dividing the squared derivatives by the probabilities in Eq.~\eqref{eq:QFI} therefore makes the contributions from the weakly populated neighbouring states dominant, as detailed in Appendix~\ref{app:pop}. Equation~\eqref{eq:Flead} shows the two features that determine the QFI. The factor $\chi_\FT^2$ describes the temperature dependence of the transitions, while the bracketed terms account for their probabilities through $\Kt_{\pm,\FT}$ and the initial occupation $n_0$. Finite-time effects change both the transition probabilities and their temperature dependence. Therefore, an increase in transition probabilities alone does not guarantee better temperature estimation. We next determine how different reservoir frequencies affect these probabilities and their temperature dependence.

Substituting Eqs.~\eqref{eq:gm} and \eqref{eq:gp} into Eq.~\eqref{eq:K} and performing the time integral gives
\begin{align}
\Kt_{-,\FT}(T,t)&=\lambda^2t^2\int_0^\infty\!\dd\omega\,J(\omega)\left[n_T(\omega)+1\right]\Phi_t(\omega),\nonumber\\
\Kt_{+,\FT}(T,t)&=\lambda^2t^2\int_0^\infty\!\dd\omega\,J(\omega)n_T(\omega)\Phi_t(\omega), \label{eq:Kint}
\end{align}
where the finite-time filter is
\begin{equation}
\Phi_t(\omega)=\sinc^2\left[\frac{(\omega-\omega_0)t}{2}\right]. \label{eq:filter}
\end{equation}
These integrals resolve the transition coefficients into contributions from different reservoir frequencies. The spectrum $J(\omega)$ describes the distribution of coupling strengths, the occupation factors distinguish upward and downward transitions, and $\Phi_t(\omega)$ weights their contributions during the interrogation. The central lobe of the filter has a width proportional to $1/t$, so longer interaction times make the frequency weighting increasingly selective around $\omega_0$. 


\begin{figure*}[t]
\centering
  \includegraphics[width=\linewidth]{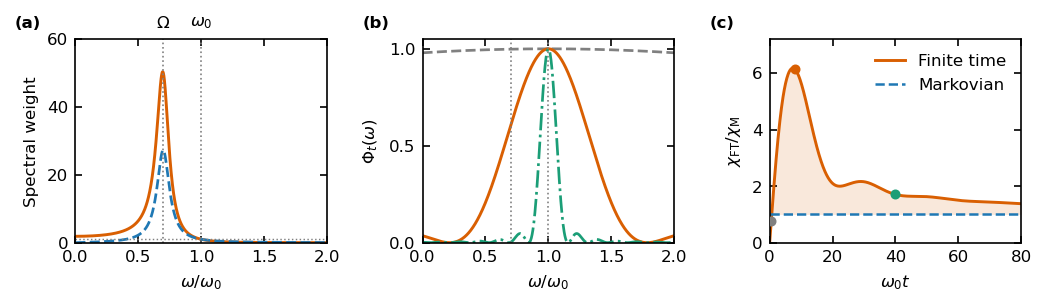}
\caption{Spectral origin of the enhanced temperature response. (a) Reservoir spectrum $J(\omega)$ (blue dashed) and thermal spectral weight $W_T(\omega)$ (orange solid), defined in Eqs.~\eqref{eq:Jform} and \eqref{eq:WT}, respectively, each normalized by its value at $\omega_0$. Vertical dotted lines mark $\Omega$ and $\omega_0$. (b) The finite-time filter in Eq.~\eqref{eq:filter} at $\omega_0t=0.5$ (gray dashed), 8 (orange solid), and 40 (green dash-dotted). Vertical dotted lines indicate the same frequencies as in (a). (c) The response ratio $\Rchi$ in Eq.~\eqref{eq:Rchi} (orange solid). The blue dashed line marks unity, and shading indicates a stronger temperature response than the Markovian benchmark. Gray, orange, and green markers correspond to the respective interrogation times in (b). Parameters are $k_BT/(\hbar\omega_0)=0.40$, $\Omega/\omega_0=0.70$, $\Gamma/\omega_0=0.12$, and $\kappa/\omega_0=1$.}
\label{fig:1}
\end{figure*}

To determine the temperature dependence of these transitions, we differentiate $\Kt_{+,\FT}$ in Eq.~\eqref{eq:Kint} with respect to $T$, holding the interrogation time, probe frequency, and reservoir spectral and coupling parameters fixed. Using Eq.~\eqref{eq:chi} then gives
\begin{equation}
\chi_\FT(T,t)=\lambda^2t^2\int_0^\infty\!\dd\omega\,W_T(\omega)\Phi_t(\omega), \label{eq:chiint}
\end{equation}
where
\begin{equation}
W_T(\omega)=J(\omega)\,\partial_Tn_T(\omega). \label{eq:WT}
\end{equation}
The thermal spectral weight $W_T(\omega)$ describes how strongly reservoir modes at frequency $\omega$ are coupled to the probe and how their occupation changes with temperature. It identifies frequencies that can contribute strongly to the temperature dependence of the probe populations when the filter gives them appreciable weight. The spectrum $J(\omega)$ alone does not contain this information about the temperature response.

We compare this response with the Markovian benchmark at the same temperature and interrogation time. Integrating the constant rates in Eq.~\eqref{eq:markov} and differentiating the resulting integrated coefficients with respect to temperature gives
\begin{align}
\Kt_{-,\Mk}&=\gamma_0t(\bar n+1),\qquad \Kt_{+,\Mk}=\gamma_0t\bar n,\nonumber\\
\chi_\Mk&=\gamma_0t\,\partial_T\bar n=2\pi\lambda^2t\,W_T(\omega_0). \label{eq:chiM}
\end{align}
Here, $\chi_\Mk$ is the temperature derivative of the Markovian integrated transition coefficients. The Markovian response samples the thermal spectral weight only at the probe frequency $\omega_0$. Dividing Eq.~\eqref{eq:chiint} by this response $\chi_\Mk$ gives
\begin{equation}
\Rchi(T,t)=\frac{\chi_\FT}{\chi_\Mk}=\frac{t}{2\pi W_T(\omega_0)}\int_0^\infty\!\dd\omega\,W_T(\omega)\Phi_t(\omega). \label{eq:Rchi}
\end{equation}
The ratio $\Rchi$ compares the temperature derivatives of the integrated transition coefficients $\Kt_{\pm,\FT}(T,t)$. When $\Rchi>1$, the leading neighbouring-state probabilities change more strongly with temperature than in the Markovian benchmark. The ratio does not depend on the initial occupation $n_0$. For the spectrum considered here, the ratio is well defined at $T>0$ and $t>0$.

To determine whether this stronger response produces a QFI advantage, we must also account for the transition probabilities. Applying Eq.~\eqref{eq:Flead} to both evolutions and taking their ratio gives
\begin{equation}
\RQ^{\rm lead}=\frac{\F^{\rm lead}_\FT}{\F^{\rm lead}_\Mk}=\Rchi^2\RK, \label{eq:factor}
\end{equation}
where
\begin{equation}
\RK=\frac{n_0/\Kt_{-,\FT}+(n_0+1)/\Kt_{+,\FT}}{n_0/\Kt_{-,\Mk}+(n_0+1)/\Kt_{+,\Mk}}. \label{eq:RK}
\end{equation}
Finite-time sampling of off-resonant reservoir modes changes both the integrated transition coefficients $\Kt_\pm$ and their temperature derivatives. The ratio $\Rchi$ compares these derivatives with the Markovian benchmark, whereas $\RK$ compares the inverse-$\Kt_\pm$ terms in the leading QFI given in Eq.~\eqref{eq:Flead}. If both integrated coefficients exceed their Markovian counterparts, $\RK<1$, so their temperature derivatives must increase sufficiently to outweigh this reduction. Equation~\eqref{eq:factor} expresses this balance through the condition $\Rchi^2\RK>1$, showing why increased transition probabilities alone do not establish a thermometric advantage. This enhancement criterion applies when the leading population expressions and their temperature derivatives accurately describe both evolutions. In particular, the probabilities of leaving the initial state must remain small, $\varepsilon_\FT\ll1$ and $\varepsilon_\Mk\ll1$, where $\varepsilon_\Mk=n_0\Kt_{-,\Mk}+(n_0+1)\Kt_{+,\Mk}$. Appendix~\ref{app:pop} discusses these validity conditions, and the following subsection tests the analytical prediction against numerical results.

Figure~\ref{fig:1} explains how off-resonant reservoir modes can strengthen the temperature dependence of the transition probabilities, increasing the sensitivity factor $\Rchi^2$ in Eq.~\eqref{eq:factor}. Figure~\ref{fig:1}(a) compares $J(\omega)$ and $W_T(\omega)$, defined in Eqs.~\eqref{eq:Jform} and \eqref{eq:WT}, respectively, each normalized by its value at the probe frequency. Their different profiles show why coupling strength alone does not determine a mode's contribution to temperature sensing. The temperature dependence of its occupation also enters through $W_T(\omega)$. For the chosen spectrum and temperature, $W_T(\omega)/W_T(\omega_0)$ is substantially larger than unity over a region below $\omega_0$. Sampling this region can therefore strengthen the temperature dependence of the transitions, whereas the Markovian response samples only the thermal spectral weight at resonance. Figure~\ref{fig:1}(b) shows how the finite-time filter in Eq.~\eqref{eq:filter} weights these off-resonant modes. At short interrogation times, its broad central lobe includes the region of elevated thermal weight identified in Fig.~\ref{fig:1}(a). These modes consequently contribute to $\chi_\FT$ through Eq.~\eqref{eq:chiint}. As the interaction time increases, the central lobe narrows around $\omega_0$, making the frequency weighting increasingly selective near resonance. However, the response also contains the prefactor $t^2$. Access to informative modes is therefore insufficient by itself, since their effect on the probe must develop during the interaction.

Figure~\ref{fig:1}(c) shows the resulting response ratio $\Rchi$. At very short times, $\Phi_t(\omega)\to1$, so Eq.~\eqref{eq:chiint} gives $\chi_\FT\propto t^2$ for the spectrum considered here. The Markovian response grows linearly with time, giving $\Rchi\propto t$. Thus, the broad short-time filter does not immediately produce a stronger temperature response. At intermediate times, the population response has developed while the filter still gives appreciable weight to the off-resonant region of elevated thermal weight. Their combined effect yields $\chi_\FT>\chi_\Mk$. This increases the sensitivity factor $\Rchi^2$, providing the mechanism through which finite-time sampling can enhance the leading QFI. At longer times, the frequency weighting becomes increasingly concentrated around $\omega_0$. The filter and integrated-response limits in Eqs.~\eqref{eq:B11} and \eqref{eq:B12} give $\Rchi\to1$. The sensitivity gain identified in Fig.~\ref{fig:1} produces a leading QFI advantage only when it outweighs any opposing change in the population factor, so that $\Rchi^2\RK>1$ in Eq.~\eqref{eq:factor}.
\begin{figure*}[t]
\centering
\includegraphics[width=\textwidth]{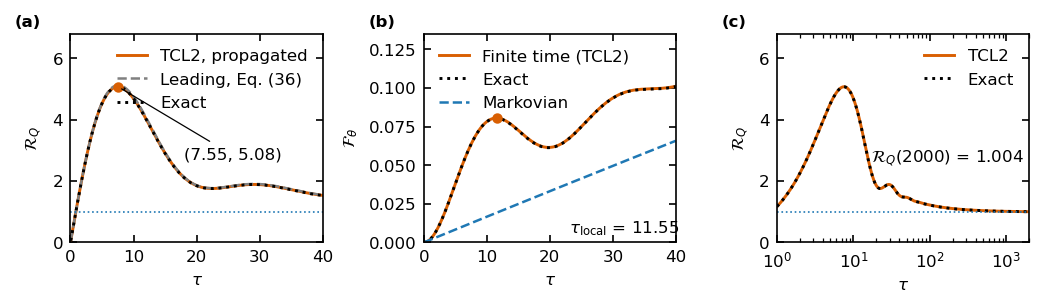}
\caption{Transient QFI enhancement and comparison with the Markovian benchmark. (a) The QFI ratio $\RQ$ in Eq.~\eqref{eq:RQ}, obtained by numerically integrating Eq.~\eqref{eq:me} and its Markovian counterpart with rates~\eqref{eq:markov} (orange solid), compared with the analytical prediction in Eq.~\eqref{eq:factor} (gray dashed) and with the exact solution of Sec.~\ref{sec:exact} (black dotted). The marker identifies the maximum relative enhancement. The analytical curve is displayed only where $\varepsilon_\FT,\varepsilon_\Mk<0.05$. (b) Absolute QFIs $\F_{\theta,X}$ for finite-time (orange solid), exact (black dotted), and Markovian (blue dashed) dynamics. The marker identifies the first local maximum of the finite-time QFI. (c) The numerically calculated QFI ratio over an extended interval, shown on a logarithmic time axis, for the TCL2 (orange solid) and exact (black dotted) dynamics. Blue dotted lines in panels (a) and (c) mark unity. Parameters are $\theta=0.40$, $\Omega/\omega_0=0.70$, $\Gamma/\omega_0=0.12$, $\kappa/\omega_0=1$, $\lambda=0.0075$, and $n_0=4$.}
\label{fig:2}
\end{figure*}
\subsection{Numerical results and interpretation}

We now determine whether the stronger temperature response identified in Fig.~\ref{fig:1} outweighs the accompanying changes in transition
probabilities and produces a QFI advantage. We propagate the populations and their temperature derivatives using Eq.~\eqref{eq:B1} and its
Markovian counterpart, then evaluate the QFI from Eq.~\eqref{eq:QFI}. This calculation includes successive transitions within the
second-order master equation without imposing the additional weak-transition expansion. To assess the accuracy of the TCL2 description, we also
propagate the populations and their temperature derivatives using the exact time-local rates in Eq.~\eqref{eq:exrates} and evaluate the corresponding QFI.
The numerical methods and convergence checks are presented in Appendix~\ref{app:num}. We express the numerical results in terms of
\begin{equation}
\theta=\frac{k_BT}{\hbar\omega_0},\qquad\tau=\omega_0t, \label{eq:dimless}
\end{equation}
where $\theta$ and $\tau$ are the dimensionless temperature and interrogation time. Applying the chain rule to Eq.~\eqref{eq:QFI} gives
\begin{equation}
\F_X(T,t)=\left(\frac{k_B}{\hbar\omega_0}\right)^2\F_{\theta,X}(\theta,\tau),
\end{equation}
where $\F_{\theta,X}$ is the QFI for estimating $\theta$, with $X\in\{\FT,\Mk,\ex\}$. The common conversion factor cancels from $\RQ$. Both evolutions use the same initial Fock state and physical parameters, specified in Fig.~\ref{fig:2}. The spectrum and temperature match those in Fig.~\ref{fig:1}, connecting the spectral mechanism directly to its numerical test.
\begin{figure*}[t!]
\centering
\includegraphics[width=0.92\textwidth]{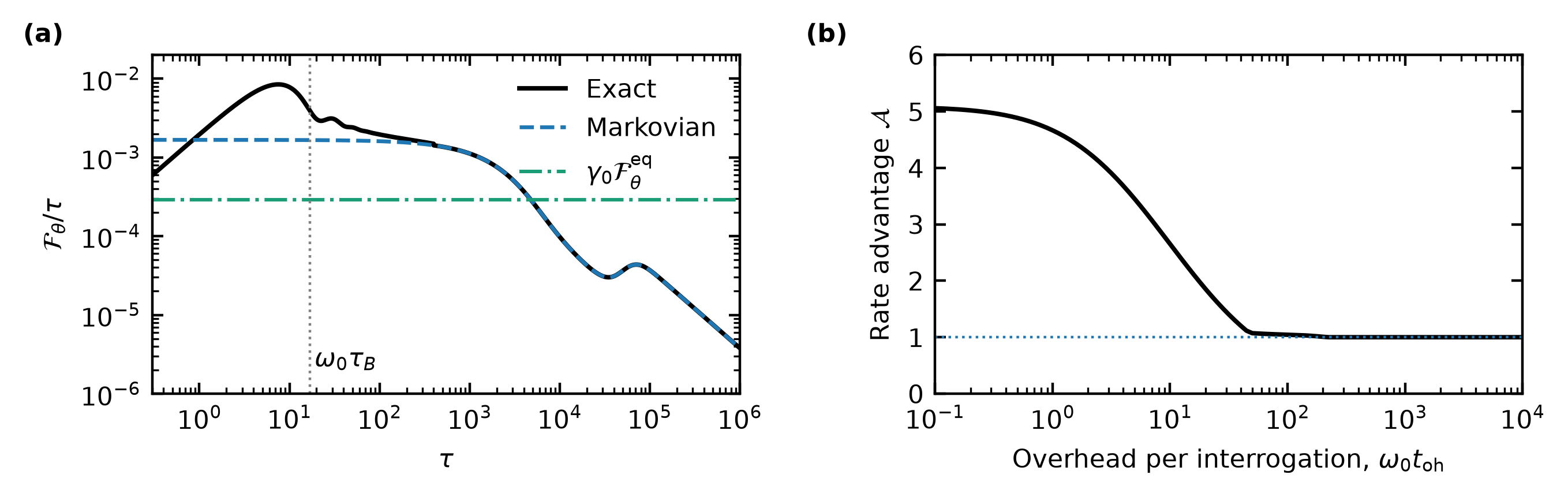}
\caption{Information per unit interrogation time. (a) Rate $\F_\theta/\tau$ for the exact dynamics (black solid) and the Markovian description (blue dashed), with $t_{\rm oh}=0$. Beyond $\tau=400$ the exact curve uses the Markovian populations, since $\RQ$ is within $0.4\%$ of unity there [Fig.~\ref{fig:2}(c)]. The green dash-dotted line marks $\gamma_0\F_\theta^{\rm eq}$, an upper bound on the equilibrium rate. The vertical dotted line marks the reservoir memory time $\omega_0\tau_B\simeq17$. (b) Rate advantage $\mathcal{A}$ in Eq.~\eqref{eq:adv} versus the overhead per interrogation. The blue dotted line marks unity. Parameters are those of Fig.~\ref{fig:2}.}
\label{fig:rate}
\end{figure*}
Figure~\ref{fig:2}(a) shows that the finite-time QFI is approximately five times the Markovian prediction at the same interrogation time. The leading expression in Eq.~\eqref{eq:factor} closely follows the numerically calculated ratio. The exact solution coincides with the TCL2 result. Their QFIs differ by less than $0.3\%$ over $0<\tau\le60$, so the TCL2 truncation does not affect our conclusions. The maximum lies at $\tau\simeq7.5$, well inside the reservoir memory time $\omega_0\tau_B\simeq17$. 
In this interval, the probability of leaving the initial state is small in both evolutions, and the individual leading QFIs agree with their numerical counterparts. These checks support the weak-transition description of the enhancement explained in Sec.~\ref{sec:mech}. Near the ratio maximum, $\Rchi^2>1$, while $\RK<1$, and their product exceeds unity. Thus, the gain in temperature sensitivity outweighs the opposing effect of the change in transition probabilities. This balance confirms the analytical mechanism developed in Sec.~\ref{sec:mech}, with the sensitivity gain arising from the off-resonant contributions identified in Fig.~\ref{fig:1}.

Figure~\ref{fig:2}(b) shows the absolute QFI per interrogation. The finite-time QFI reaches its first local maximum later than the maximum relative enhancement in Fig.~\ref{fig:2}(a). These times differ because $\RQ$ compares two evolving QFIs, whereas $\F_{\theta,\FT}$ measures the information carried by the finite-time probe state itself. The marker identifies the first local maximum and does not establish a global optimum. A large relative enhancement, therefore, need not coincide with the largest absolute information. Figure~\ref{fig:2}(c) extends the comparison to longer times and shows the QFI ratio approaching unity. This behavior is consistent with the increasingly resonance-selective response discussed in Sec.~\ref{sec:mech}. The curve is calculated from numerically propagated populations throughout the displayed interval. Together, Figs.~\ref{fig:1} and \ref{fig:2} explain the origin of the increased temperature sensitivity and demonstrate that it produces a transient QFI advantage. The scans in Appendix~\ref{app:num} show that the advantage persists across the investigated linewidths at the baseline reservoir frequency, while also identifying settings without enhancement within the sampled time interval.

\subsection{Information per unit time}\label{sec:rate}

The ratio $\RQ$ compares two descriptions of the same experiment at a fixed interrogation time. To assess the operational benefit of the finite-time response, we consider the temperature information acquired within a fixed total experimental time. For independent repetitions of duration $t+t_{\rm oh}$, where $t_{\rm oh}$ accounts for state preparation and readout, the number of measurements is $m\simeq t_{\rm tot}/(t+t_{\rm oh})$. Equation~\eqref{eq:QCRB} then gives $\Delta^2T\ge[t_{\rm tot}\F_X/(t+t_{\rm oh})]^{-1}$. The information rate $\F_X/(t+t_{\rm oh})$ therefore determines the precision bound for a given time budget.

Figure~\ref{fig:rate}(a) compares the information rates for negligible overhead. Under the exact dynamics, the rate reaches a maximum within the reservoir memory time. At this interrogation time, the probe has acquired appreciable temperature information while the interaction remains short enough to allow frequent repetition. The Markovian description instead predicts a nearly constant rate at short times, reflecting the linear growth of its QFI. The transient peak is about five times this Markovian rate. Because the latter varies little over the weak-transition regime, the rate advantage closely follows the same-time QFI enhancement in Fig.~\ref{fig:2}(a).

The benefit of short interrogations is also evident when comparing different interaction durations under the exact dynamics. For interrogations longer than three reservoir memory times, the information acquired per unit time remains below the transient peak over the investigated interval. Although longer interactions can increase the information obtained from an individual measurement, they allow fewer repetitions within the total time budget. Equilibrium thermometry incurs the additional cost of waiting for the probe to thermalize. The equilibrium QFI, $\F_\theta^{\rm eq}=\bar n(\bar n+1)/\theta^4$, must therefore be considered together with the relaxation timescale $1/\gamma_0$. Taking this timescale as an optimistic equilibration-cycle duration gives the dimensionless rate estimate $(\gamma_0/\omega_0)\F_\theta^{\rm eq}$ shown in Fig.~\ref{fig:rate}(a). The transient optimum is about 30 times higher than this estimate.
\begin{figure*}[t]
\centering
\includegraphics[width=0.92\textwidth]{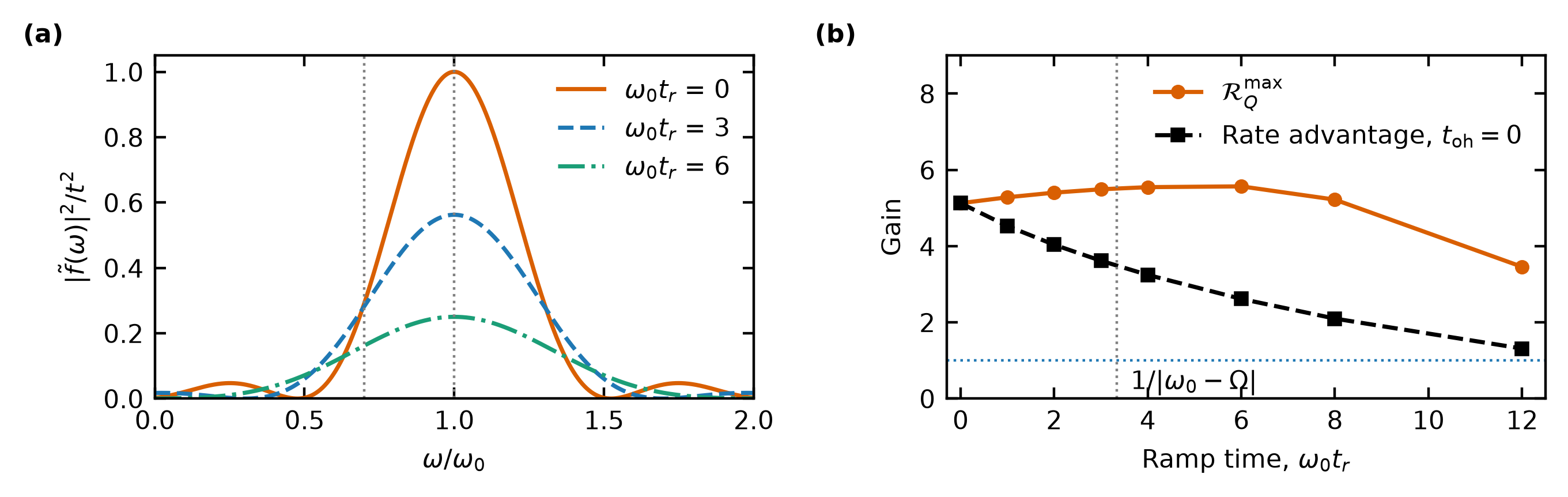}
\caption{Effect of a finite switching time. (a) Normalized filter $|\tilde f(\omega)|^2/t^2$ from Eq.~\eqref{eq:ftilde} at $\omega_0t=12$ for ramp times $\omega_0t_r=0$ (orange solid), 3 (blue dashed), and 6 (green dash-dotted). Vertical dotted lines mark $\Omega$ and $\omega_0$. (b) Maximum QFI ratio against the Markovian benchmark with the same envelope (orange circles) and rate advantage over the best Markovian rate at $t_{\rm oh}=0$ (black squares), both maximized over $0<\omega_0t\le80$. The vertical dotted line marks $\omega_0t_r=1/|\omega_0-\Omega|$, and the horizontal dotted line marks unity. Results use the leading-order expressions. Other parameters are those of Fig.~\ref{fig:2}.}
\label{fig:switch}
\end{figure*}
Preparation and readout overhead reduce the benefit of frequent repetition. To quantify this effect, we compare the optimized rates,
\begin{equation}
\mathcal{A}(t_{\rm oh})=\frac{\max_t\F_{\theta,\ex}(t)/(t+t_{\rm oh})}{\max_t\F_{\theta,\Mk}(t)/(t+t_{\rm oh})}.
\label{eq:adv}
\end{equation}
Figure~\ref{fig:rate}(b) shows that small overhead preserves most of the transient advantage. As the overhead increases, short interrogations devote a larger fraction of each cycle to preparation and readout. Longer interactions then become more competitive, and the rate advantage decreases. For the parameters studied, retaining a substantial improvement requires preparation and readout within a few tens of inverse probe frequencies. This condition connects the predicted advantage to the duration of the complete measurement cycle. Lower-frequency probes allow a longer overhead in physical time for the same dimensionless constraint, provided that their preparation and readout can be completed on that timescale.

\subsection{Finite switching time}\label{sec:switch}

So far the coupling has switched on and off instantaneously. The filter in Eq.~\eqref{eq:filter} is the Fourier transform of this rectangular window, and its slowly decaying side lobes reach far from $\omega_0$. A real coupler switches in a finite time. We therefore replace $\lambda$ in Eq.~\eqref{eq:HI} by $\lambda f(s)$ for $0\le s\le t$. The envelope $f(s)$ rises as $\sin^2[\pi s/(2t_r)]$ over a ramp time $t_r$, stays at unity, and falls symmetrically at the end. To leading order in $\lambda$, the switching enters only through the filter. Appendix~\ref{app:switch} shows that $t^2\Phi_t(\omega)$ in Eqs.~\eqref{eq:Kint} and \eqref{eq:chiint} becomes $|\tilde f(\omega)|^2$, with
\begin{equation}
\tilde f(\omega)=\int_0^t\!\dd s\,f(s)\,e^{i(\omega_0-\omega)s}. \label{eq:ftilde}
\end{equation}
The Markovian benchmark uses the same envelope, which replaces $t$ by $t_{\rm eff}=\int_0^tf^2(s)\,\dd s$ in Eq.~\eqref{eq:chiM}.
\begin{figure}[b]
\centering
\includegraphics[width=\columnwidth]{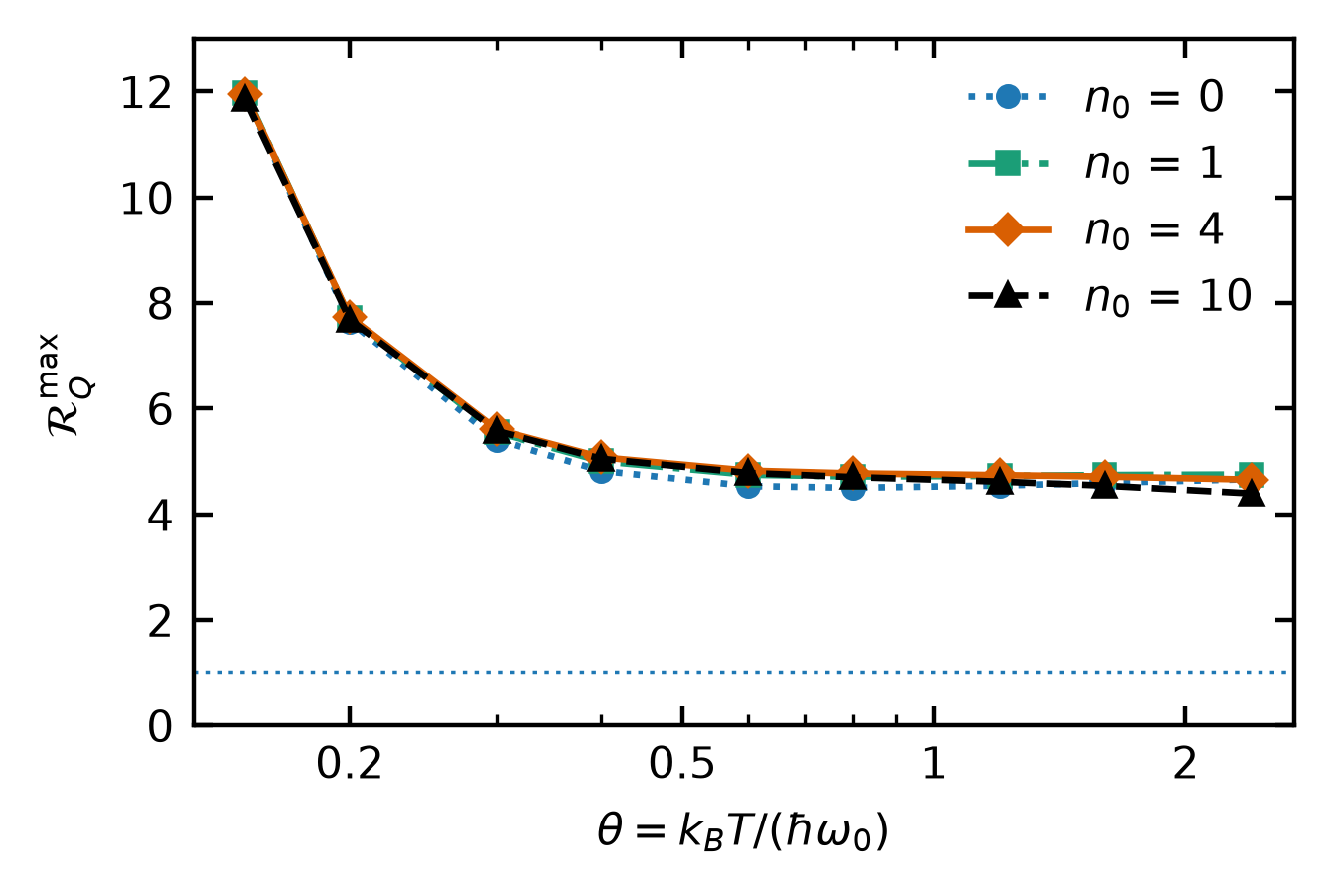}
\caption{Temperature and initial-state dependence of the transient QFI enhancement. The maximum propagated ratio $\RQ^{\max}$ over $0<\tau\le60$ is shown versus $\theta$ for $n_0=0$ (blue circles), 1 (green squares), 4 (orange diamonds), and 10 (black triangles). The dotted line marks unity. Other parameters are those of Fig.~\ref{fig:2}.}
\label{fig:scan}
\end{figure}

Figure~\ref{fig:switch}(a) shows that ramps suppress the side lobes and lower the central lobe at fixed $t$. The relative gain survives. Figure~\ref{fig:switch}(b) shows that $\RQ^{\max}$ stays between 5.1 and 5.6 for $\omega_0t_r\le8$ and falls to 3.5 at $\omega_0t_r=12$. The informative modes near $\Omega$ lie inside the central lobe of the filter, so the side lobes contribute little. The gain is determined by the finite interaction duration, without requiring abrupt switching.

Ramps require additional time, extending the interrogation without providing full-strength coupling and thereby reducing the information rate. With $t_{\rm oh}=0$, the rate advantage over the best Markovian rate drops from 5.1 for instantaneous switching to 3.6 at $\omega_0t_r=3$, 2.6 at $\omega_0t_r=6$, and 1.3 at $\omega_0t_r=12$. Retaining most of the rate advantage therefore requires ramps shorter than about $1/|\omega_0-\Omega|$, which equals $3.3/\omega_0$ here. All results in this subsection satisfy $\varepsilon_\FT<0.012$ at the ratio maxima, which keeps them inside the weak-transition regime.

\subsection{Temperature and initial-state dependence}\label{sec:scan}

The gain persists over the temperature range we examine. Figure~\ref{fig:scan} shows the maximum propagated ratio over $0<\tau\le60$ for $0.15\le\theta\le2.4$. At $n_0=4$ it rises from 4.7 at high temperature to 7.7 at $\theta=0.2$ and 11.9 at $\theta=0.15$. The enhancement is strongest for cold reservoirs. In this regime $\partial_Tn_T(\omega)$ grows rapidly toward low frequencies, so the modes below $\omega_0$ carry much more thermal weight than the resonant mode. The maximizing time stays between $\tau\simeq6.9$ and 7.7 at all temperatures, which reflects the fixed spectral detuning.

The initial number state has little effect on the enhancement. At $\theta=0.4$, the maximum ratio is 4.8 for the vacuum, 5.0 for $n_0=1$, 5.1 for $n_0=4$, and 5.0 for $n_0=10$. The curves for different $n_0$ nearly coincide at all temperatures. This follows from  Eq.~\eqref{eq:factor}: the sensitivity factor $\Rchi$ is independent of $n_0$, while $n_0$ only reweights the two inverse-$\Kt_\pm$ terms in $\RK$. Thus, the finite-time gain is primarily determined by the reservoir response rather than the initial number state. Fock states still change the absolute information, which is the role studied in Refs.~\cite{UllahNaseemMustecaplioglu2025,gsh7-r7ms}.

\section{Conclusion}
\label{sec:conclusion}

We have shown that finite-time reservoir response can enhance transient thermometry with a harmonic-oscillator probe prepared in a
number state. Off-resonant reservoir modes can strengthen the temperature dependence of the probe populations and thereby increase the QFI above the
Markovian prediction. Our analysis identifies which reservoir frequencies contribute most
strongly to the temperature dependence of the probe populations. When the probability of leaving the initial state is small, the
analytical criterion determines whether this increased sensitivity outweighs the accompanying changes in transition probabilities.
Comparison with the exact dynamics confirms the transient enhancement predicted by the second-order master equation.

The operational significance of this enhancement follows from the temperature information acquired per unit time. For the parameters
studied, the exact dynamics yield an optimal interrogation within the reservoir memory time, with an information rate exceeding both the
Markovian benchmark and that available from longer interrogations. Preparation and readout overhead reduce this advantage, making the
duration of the full experimental cycle an essential consideration. The enhancement survives smooth switching because the informative
reservoir modes remain within the central lobe of the finite-time filter, although longer ramps reduce the information rate.
Across the investigated range, the relative QFI enhancement increases at lower temperatures and depends weakly on the initial occupation.
Ideal excitation-number measurements attain the QFI because the probe remains diagonal in a temperature-independent number basis.

An experimental test would combine repeated number-state preparation, a controlled interaction duration, and measurements of the excitation
populations. The reservoir temperature could be inferred by fitting these populations to the predicted dynamics. Measurements at nearby
calibrated temperatures would also allow the population derivatives and the corresponding Fisher information to be estimated at different
interrogation times. Trapped-ion experiments have demonstrated heating and decay of motional Fock states in engineered
reservoirs~\cite{PhysRevA.62.053807}, while superconducting resonators support deterministic oscillator-state preparation and Wigner
tomography~\cite{Hofheinz2009}. These capabilities provide a starting point for an implementation with a characterized thermal reservoir
and suitable off-resonant spectral weight. Observing the predicted information-rate advantage would further require sufficiently rapid
coupling control, preparation, and readout. These findings provide a basis for choosing interrogation times that
improve the precision of transient thermometry under experimental timing constraints.

\section*{Acknowledgments}

M. Tahir Naseem acknowledges support from the National Center for Quantum Computing (NCQC), Pakistan. A. U. and \"O. E. M. acknowledge financial support from the Scientific and Technological Research Council of T\"urkiye (T\"UB\.ITAK) under Project Grant No. 123F150.

\section*{Data availability}
The data that support the findings of this article are available
from the authors upon reasonable request.
\appendix
\renewcommand{\thesection}{\Alph{section}}
\renewcommand{\theequation}{\thesection\arabic{equation}}
\titleformat{\section}{\normalfont\small\bfseries\centering}{Appendix \thesection:}{0.5em}{}

\section{Derivation of the finite-time master equation}\label{app:me}
\setcounter{equation}{0}

We derive the finite-time master equation to second order in the probe--reservoir coupling. Transforming the interaction Hamiltonian in Eq.~\eqref{eq:HI} with the free Hamiltonian $H_S+H_B$ gives
\begin{equation}
\tilde H_I(t)=\hbar\lambda\left[a^\dagger e^{i\omega_0t}B(t)+ae^{-i\omega_0t}B^\dagger(t)\right],
\end{equation}
where the tilde denotes the interaction picture and
\begin{equation}
B(t)=\sum_kg_kb_ke^{-i\omega_kt}.
\end{equation}
For the initially factorized state in Eq.~\eqref{eq:init}, the second-order time-convolutionless expansion gives
\begin{equation}
\frac{\dd\tilde\rho(t)}{\dd t}=-\frac{1}{\hbar^2}\int_0^t\!\dd s\,\mathrm{Tr}_B\left[\tilde H_I(t),\left[\tilde H_I(s),\tilde\rho(t)\otimes\rho_B(T)\right]\right]. \label{eq:A3}
\end{equation}
Here, $\tilde\rho(t)$ is the reduced probe state, and $\rho_B(T)$ is the initial thermal reservoir state. We suppress the temperature argument of the reduced state throughout this Appendix. In Eq.~\eqref{eq:A3}, the first-order contribution vanishes because the thermal reservoir has $\langle b_k\rangle_B=\langle b_k^\dagger\rangle_B=0$, which implies $\mathrm{Tr}_B[\tilde H_I(t)\rho_B(T)]=0$. Thus, the leading reservoir-induced change in the reduced state is second order in $\lambda$. Equation~\eqref{eq:A3} specifies the truncated generator without assuming that the exact joint state remains factorized during the evolution. The finite integration interval is retained.

To evaluate the bath trace in Eq.~\eqref{eq:A3}, we first expand the double commutator,
\begin{equation}
[H_t,[H_s,R]]=H_tH_sR-H_tRH_s-H_sRH_t+RH_sH_t,
\end{equation}
where $H_t=\tilde H_I(t)$, $H_s=\tilde H_I(s)$, and $R=\tilde\rho(t)\otimes\rho_B(T)$. Substituting $\tilde H_I$ into these terms produces products containing two reservoir operators. Their thermal traces follow from $\langle b_kb_\ell^\dagger\rangle_B=\delta_{k\ell}[n_T(\omega_k)+1]$, $\langle b_k^\dagger b_\ell\rangle_B=\delta_{k\ell}n_T(\omega_k)$, and $\langle b_kb_\ell\rangle_B=\langle b_k^\dagger b_\ell^\dagger\rangle_B=0$. Here, $\langle O\rangle_B=\mathrm{Tr}_B[O\rho_B(T)]$, and $n_T(\omega)$ is the Bose occupation. Using the bath operator $B(t)$ and these thermal averages, then converting the mode sums to frequency integrals using Eq.~\eqref{eq:J}, gives the two-time bath correlations
\begin{align}
C_-(u)&=\langle B(u)B^\dagger(0)\rangle_B=\int_0^\infty\!\dd\omega\,J(\omega)[n_T(\omega)+1]e^{-i\omega u},\\
C_+(u)&=\langle B^\dagger(u)B(0)\rangle_B=\int_0^\infty\!\dd\omega\,J(\omega)n_T(\omega)e^{i\omega u}.
\end{align}
Here, $u=t-s$ is the time difference. Thermal stationarity makes the correlations depend only on $u$ and gives $C_\pm(-u)=C_\pm^*(u)$.

Using these correlations to evaluate the four bath traces and substituting the result into Eq.~\eqref{eq:A3} yields
\begin{align}
\frac{\dd\tilde\rho}{\dd t}={}&G_-\left(a\tilde\rho a^\dagger-a^\dagger a\tilde\rho\right)+G_-^*\left(a\tilde\rho a^\dagger-\tilde\rho a^\dagger a\right)\nonumber\\
&+G_+\left(a^\dagger\tilde\rho a-aa^\dagger\tilde\rho\right)+G_+^*\left(a^\dagger\tilde\rho a-\tilde\rho aa^\dagger\right),
\end{align}
where $\tilde\rho=\tilde\rho(t)$ and $G_\pm=G_{\pm,\FT}(T,t)$ abbreviate the response coefficients obtained after changing variables from $s$ to $u$,
\begin{align}
G_{-,\FT}(T,t)&=\lambda^2\int_0^t\!\dd u\,e^{i\omega_0u}C_-(u),\nonumber\\
G_{+,\FT}(T,t)&=\lambda^2\int_0^t\!\dd u\,e^{-i\omega_0u}C_+(u). \label{eq:A8}
\end{align}
The vanishing anomalous bath correlations eliminate terms containing two probe creation or two probe annihilation operators. For the excitation-exchange interaction in Eq.~\eqref{eq:HI}, no additional secular averaging is required. Retaining the upper limit $t$ in Eq.~\eqref{eq:A8} preserves the transient reservoir response. Extending these correlation-time integrals to infinity gives the Markovian coefficients when the corresponding limits exist.

Substituting the correlations into Eq.~\eqref{eq:A8} and performing the finite-time integration gives
\begin{align}
G_{-,\FT}(T,t)&=\lambda^2\int_0^\infty\!\dd\omega\,J(\omega)[n_T(\omega)+1]\,\Lambda_+(\Delta,t),\nonumber\\
G_{+,\FT}(T,t)&=\lambda^2\int_0^\infty\!\dd\omega\,J(\omega)n_T(\omega)\,\Lambda_-(\Delta,t),
\end{align}
where
\begin{equation}
\Lambda_\pm(\Delta,t)=\frac{\sin(\Delta t)}{\Delta}\pm i\frac{1-\cos(\Delta t)}{\Delta}. \label{eq:A10}
\end{equation}
Here, $\Delta=\omega_0-\omega$ is the probe--mode detuning. At resonance, the continuous limit is $\Lambda_\pm(0,t)=t$, so the finite-time kernels are regular. We denote these kernels by $\Lambda_\pm$ to distinguish them from the integrated coefficients $\Kt_{\pm,\FT}$ in Eq.~\eqref{eq:K}.

Separating the real and imaginary contributions and returning to the Schr\"odinger picture gives Eq.~\eqref{eq:me}, with $\gamma_{\pm,\FT}=2\,\mathrm{Re}\,G_{\pm,\FT}$. The Lamb-shift Hamiltonian is
\begin{equation}
H_{LS}(T,t)=\hbar S_{-,\FT}(T,t)a^\dagger a+\hbar S_{+,\FT}(T,t)aa^\dagger, \label{eq:HLS}
\end{equation}
where $S_{\pm,\FT}=\mathrm{Im}\,G_{\pm,\FT}$ are the frequency-shift coefficients. Taking the imaginary parts, using Eq.~\eqref{eq:A10}, gives
\begin{align}
S_{-,\FT}(T,t)&=\lambda^2\int_0^\infty\!\dd\omega\,J(\omega)[n_T(\omega)+1]\frac{1-\cos(\Delta t)}{\Delta},\nonumber\\
S_{+,\FT}(T,t)&=-\lambda^2\int_0^\infty\!\dd\omega\,J(\omega)n_T(\omega)\frac{1-\cos(\Delta t)}{\Delta}.
\end{align}
The opposite signs follow from the imaginary parts of $\Lambda_+$ and $\Lambda_-$. Taking twice the real parts of the same response functions gives
\begin{align}
\gamma_{-,\FT}(T,t)&=2\lambda^2\int_0^\infty\!\dd\omega\,J(\omega)[n_T(\omega)+1]\frac{\sin(\Delta t)}{\Delta},\nonumber\\
\gamma_{+,\FT}(T,t)&=2\lambda^2\int_0^\infty\!\dd\omega\,J(\omega)n_T(\omega)\frac{\sin(\Delta t)}{\Delta}. \label{eq:A14}
\end{align}
These expressions recover Eqs.~\eqref{eq:gm} and \eqref{eq:gp}, completing the derivation of Eq.~\eqref{eq:me}.

\section{Population dynamics and analytical QFI}\label{app:pop}
\setcounter{equation}{0}

This appendix derives the population expansion and leading QFI used in Sec.~\ref{sec:mech}. It also establishes the spectral identities and limiting behavior underlying the finite-time thermal response. We take $T>0$ and hold the interrogation time and all Hamiltonian parameters fixed when differentiating with respect to temperature. Taking diagonal matrix elements of Eq.~\eqref{eq:me} in the number basis gives
\begin{align}
\dot p_n^\FT={}&\gamma_{-,\FT}\left[(n+1)p^\FT_{n+1}-np_n^\FT\right]\nonumber\\
&+\gamma_{+,\FT}\left[np^\FT_{n-1}-(n+1)p_n^\FT\right]. \label{eq:B1}
\end{align}
Here, $n\geq0$, $p^\FT_{-1}=0$, and the common arguments $(T,t)$ are suppressed. The factors $n$ and $n+1$ follow from the number-state matrix elements of $a$ and $a^\dagger$. The Hamiltonian terms do not change the populations because both the free and Lamb-shift Hamiltonians are diagonal in this basis. The Markovian population equation follows by replacing the finite-time coefficients with Eq.~\eqref{eq:markov}.

Integrating Eq.~\eqref{eq:B1} once and inserting the initial populations $p_n(0)=\delta_{n,n_0}$ on its right-hand side gives Eq.~\eqref{eq:pop}, with $\Kt_{\pm,\FT}$ and $\varepsilon_\FT$ defined in Eqs.~\eqref{eq:K} and \eqref{eq:eps}. At fixed $T$ and $t$, the transition coefficients and their integrals are of order $\lambda^2$. A further iteration therefore introduces corrections of order $\lambda^4$, including populations reached through successive transitions. This order counting describes an expansion of the adopted second-order master equation. It does not supply the higher-order terms omitted from its generator. To evaluate the integrated transition coefficients, we substitute Eqs.~\eqref{eq:gm} and \eqref{eq:gp} into Eq.~\eqref{eq:K}. The required time integral is
\begin{equation}
\int_0^t2s\,\sinc(\Delta s)\,\dd s=\frac{2[1-\cos(\Delta t)]}{\Delta^2}=t^2\sinc^2\!\left(\frac{\Delta t}{2}\right),
\end{equation}
where $\Delta=\omega_0-\omega$, as in Appendix~\ref{app:me}, and the expression at $\Delta=0$ is understood by continuity. Using this result in the frequency integrals gives Eq.~\eqref{eq:Kint}. Since $J(\omega)$, $n_T(\omega)$, and $\Phi_t(\omega)$ are nonnegative, the integrated coefficients $\Kt_{\pm,\FT}$ are nonnegative, even when the instantaneous coefficients change sign.

Subtracting the two expressions in Eq.~\eqref{eq:Kint} gives
\begin{equation}
\Kt_{-,\FT}-\Kt_{+,\FT}=\lambda^2t^2\int_0^\infty\!\dd\omega\,J(\omega)\Phi_t(\omega).
\end{equation}
For a temperature-independent spectrum, the right-hand side has no temperature dependence. Differentiating therefore establishes the equality of derivatives in Eq.~\eqref{eq:chi}. Differentiating the Bose occupation gives
\begin{equation}
\partial_Tn_T(\omega)=\frac{\hbar\omega}{k_BT^2}n_T(\omega)\left[n_T(\omega)+1\right].
\end{equation}
Using this derivative in Eq.~\eqref{eq:Kint} yields the thermal-response integral in Eq.~\eqref{eq:chiint}. The temperature derivatives of the leading populations follow directly from Eqs.~\eqref{eq:pop} and \eqref{eq:chi},
\begin{align}
\partial_Tp^\FT_{n_0-1}&\simeq n_0\chi_\FT,\nonumber\\
\partial_Tp^\FT_{n_0}&\simeq-(2n_0+1)\chi_\FT,\nonumber\\
\partial_Tp^\FT_{n_0+1}&\simeq(n_0+1)\chi_\FT.
\end{align}
Their sum vanishes, as required by normalization. Substituting these derivatives and the leading populations into Eq.~\eqref{eq:QFI} gives
\begin{equation}
\F^{\rm tr}_\FT=\chi_\FT^2\left[\frac{n_0}{\Kt_{-,\FT}}+\frac{n_0+1}{\Kt_{+,\FT}}\right]+\frac{(2n_0+1)^2\chi_\FT^2}{1-\varepsilon_\FT}. \label{eq:B6}
\end{equation}
Here, $\F^{\rm tr}_\FT$ is the QFI of the normalized distribution retained in Eq.~\eqref{eq:pop}, with $\varepsilon_\FT<1$. The first contribution comes from the neighbouring populations, and the second comes from the initially occupied state. Terms associated with the downward transition are omitted throughout when $n_0=0$. To compare these contributions, the Cauchy--Schwarz inequality gives
\begin{equation}
(2n_0+1)^2\leq\varepsilon_\FT\left[\frac{n_0}{\Kt_{-,\FT}}+\frac{n_0+1}{\Kt_{+,\FT}}\right]. \label{eq:B7}
\end{equation}
Using Eq.~\eqref{eq:B7} in Eq.~\eqref{eq:B6} bounds the initial-state contribution relative to the neighbouring-state contribution by $\varepsilon_\FT/(1-\varepsilon_\FT)$. It is therefore subleading when $\varepsilon_\FT\ll1$. Retaining the leading contribution gives Eq.~\eqref{eq:Flead}.

At fixed $T$ and $t$, $\Kt_{\pm,\FT}$ and $\chi_\FT$ are of order $\lambda^2$. The leading QFI is consequently of order $\lambda^2$, whereas the initial-state contribution is of order $\lambda^4$. Corrections to the neighbouring populations and the populations reached through successive transitions can also contribute at order $\lambda^4$. Thus, retaining the initial-state term alone does not give the complete QFI at that order. This counting assumes that the population expansion and its temperature derivative remain regular. Small depletion provides a useful validity check, but the accuracy of the QFI also depends on the accuracy of the temperature derivatives. For this reason, Fig.~\ref{fig:2}(a) compares the leading and propagated QFIs directly.

For the Markovian evolution, substituting Eq.~\eqref{eq:chiM} into the leading population expression gives
\begin{equation}
\F^{\rm lead}_\Mk=\gamma_0t(\partial_T\bar n)^2\left[\frac{n_0}{\bar n+1}+\frac{n_0+1}{\bar n}\right]. \label{eq:B8}
\end{equation}
This expression requires
\begin{equation}
\varepsilon_\Mk=\gamma_0t\left[n_0(\bar n+1)+(n_0+1)\bar n\right]\ll1.
\end{equation}
Taking the ratio of Eqs.~\eqref{eq:Flead} and \eqref{eq:B8} then gives the factorization in Eq.~\eqref{eq:factor}. Its interpretation applies where both population expansions and their temperature derivatives accurately describe the propagated dynamics. The short-time behavior follows from $\Phi_t(\omega)\to1$ as $t\to0$. For the spectrum in Eq.~\eqref{eq:Jform}, the frequency integrals of $J(\omega)[n_T(\omega)+1]$, $J(\omega)n_T(\omega)$, and $W_T(\omega)$ are finite. Equations~\eqref{eq:Kint} and \eqref{eq:chiint} therefore give $\Kt_{\pm,\FT}\propto t^2$ and $\chi_\FT\propto t^2$. Using these scalings in Eq.~\eqref{eq:Flead} and comparing with Eq.~\eqref{eq:B8} yields
\begin{equation}
\F^{\rm lead}_\FT\propto t^2,\qquad\F^{\rm lead}_\Mk\propto t,\qquad\RQ^{\rm lead}\propto t.
\end{equation}
Thus, broad spectral access at very short times does not produce an immediate QFI advantage. At long times, the filter satisfies the distributional limit
\begin{equation}
t\,\Phi_t(\omega)\longrightarrow2\pi\delta(\omega-\omega_0). \label{eq:B11}
\end{equation}
For the integrable spectral weights used here, which are continuous at $\omega_0>0$, applying this limit to Eqs.~\eqref{eq:Kint} and \eqref{eq:chiint} gives
\begin{equation}
\frac{\Kt_{-,\FT}}{t}\to\gamma_0(\bar n+1),\quad\frac{\Kt_{+,\FT}}{t}\to\gamma_0\bar n,\quad\frac{\chi_\FT}{t}\to\gamma_0\partial_T\bar n. \label{eq:B12}
\end{equation}
These limits establish recovery of the Markovian integrated response per unit time. They do not extend the weak-transition population expansion to arbitrarily long interrogations. The longer-time QFI comparison in Fig.~\ref{fig:2}(c) therefore uses the propagated populations.

\section{Numerical validation and spectral robustness}\label{app:num}
\setcounter{equation}{0}

To assess the dependence of the QFI enhancement on the reservoir spectrum, we propagate the populations and their temperature derivatives using Eq.~\eqref{eq:B1}, then evaluate the QFI from Eq.~\eqref{eq:QFI}. The Markovian benchmark is calculated in the same way, using the rates in Eq.~\eqref{eq:markov}. We use the adaptive DOP853 integrator with number states from 0 to 40. Frequency integrals are evaluated using composite Gauss--Legendre quadrature, with oscillatory quadrature for the extended evolution in Fig.~\ref{fig:2}(c). The frequency cutoff is $\omega/\omega_0=80$ for Fig.~\ref{fig:2} and 40 for the robustness scan. To check convergence, we increase the frequency cutoff to 80 and the number-state cutoff to 60, and refine the integration and sampling. At the tested low-frequency, near-resonant, and high-frequency settings, these changes alter the peak ratios by less than $10^{-6}$ in relative terms. We also independently validate the propagated temperature derivatives using centered finite differences. These checks establish numerical convergence at the tested settings within the adopted second-order master equation. For the exact solution of Appendix~\ref{app:exact}, we use $\delta\omega/\omega_0=0.004$ and $\omega_c/\omega_0=8$ for $\tau\le400$, and $\delta\omega/\omega_0=0.002$ for Fig.~\ref{fig:2}(c). The recurrence time then exceeds the displayed interval. Halving $\delta\omega$ and raising $\omega_c$ to $12\omega_0$ changes the maximum of $\RQ$ by less than $10^{-6}$ in relative terms. For the information-rate benchmark, the switching analysis and the temperature and initial-state scans, frequency integrals are evaluated by trapezoidal quadrature on a nonuniform grid, with spacing $0.0015\,\omega_0$ below $6\,\omega_0$ and logarithmic spacing up to the cutoff. Populations are propagated with the same DOP853 integrator. At the baseline parameters, this procedure reproduces
the maximum $\mathcal{R}_Q\simeq5.08$ at $\tau\simeq7.5$ obtained with Gauss--Legendre quadrature.

We vary the characteristic reservoir frequency $\Omega$ and linewidth parameter $\Gamma$ in Eq.~\eqref{eq:Jform}, while keeping the temperature, initial state, and remaining coupling parameters fixed. At each spectral setting, both evolutions use the same physical parameters. To compare their relative performance over a common transient interval, we define
\begin{equation}
\RQ^{\max}(\Omega,\Gamma)=\max_{0<\omega_0t\le200}\RQ(T,t;\Omega,\Gamma). \label{eq:C1}
\end{equation}
Here, $\RQ^{\max}$ is the largest propagated QFI ratio within the specified interval, with its dependence on the varied spectral parameters displayed explicitly. A maximum at the upper boundary does not establish an unrestricted optimal interrogation time.

\begin{figure}[t]
\centering
\includegraphics[width=\columnwidth]{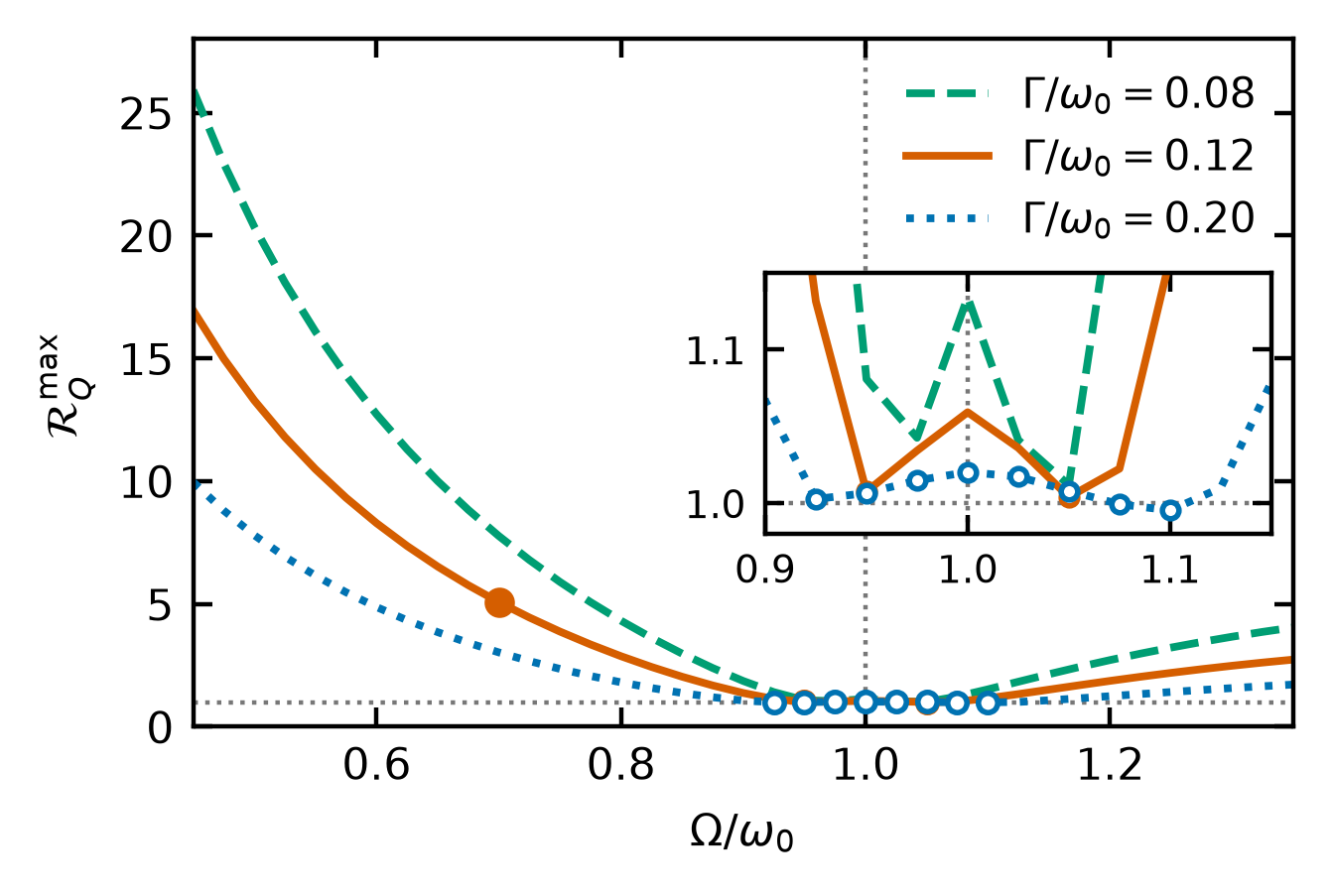}
\caption{Spectral dependence of the transient QFI enhancement. The maximum propagated ratio in Eq.~\eqref{eq:C1} is shown versus $\Omega/\omega_0$ for $\Gamma/\omega_0=0.08$ (green dashed), 0.12 (orange solid), and 0.20 (blue dotted). The scan covers $0.45\le\Omega/\omega_0\le1.35$ in steps of 0.025, with $0<\omega_0t\le200$. Lines connect the sampled spectral settings. Fixed parameters are $\theta=0.40$, $\kappa/\omega_0=1$, $\lambda=0.0075$, and $n_0=4$. The filled orange marker identifies the baseline setting used in Figs.~\ref{fig:1} and \ref{fig:2}. Open markers indicate maxima at the upper time boundary. The horizontal dotted line marks unity, the vertical dotted line marks $\Omega=\omega_0$, and the inset resolves the region near unity. All curves are calculated from propagated populations.}
\label{fig:3}
\end{figure}

Figure~\ref{fig:3} shows that the enhancement persists across the investigated linewidths at the baseline reservoir frequency used in Sec.~\ref{sec:results}. At this frequency, narrower spectra yield larger maximum QFI ratios. The relative enhancement also depends strongly on spectral position. The largest sampled ratios occur at the lower-frequency edge of the scan, while the advantage weakens near resonance for some settings. The ratios increase again at higher reservoir frequencies, showing that enhancement is not restricted to spectra with $\Omega<\omega_0$.

These trends can be interpreted through the thermal spectral weight $W_T(\omega)$ in Eq.~\eqref{eq:WT} and the finite-time response in Eq.~\eqref{eq:Rchi}. Changing $\Omega$ and $\Gamma$ changes both the thermal weight available away from resonance and the resonant contribution sampled by the Markovian benchmark. Within the weak-transition regime, Eq.~\eqref{eq:factor} connects the resulting thermal-response gain to the QFI through the accompanying population factor. A mode's frequency alone therefore does not determine its thermometric contribution. Some maximizing times in the scan lie outside this analytical regime, so the figure uses propagated QFIs throughout rather than extending the leading factorization beyond its validity. The inset in Fig.~\ref{fig:3} also identifies spectral settings for which no enhancement is found within the sampled time interval. The spectral modification therefore does not guarantee an advantage. Open markers require particular care since they indicate that the largest sampled ratio occurs at the final interrogation time. Neither the absence of enhancement nor a boundary maximum establishes the behavior outside the specified time window.

Finally, relative enhancement must be distinguished from absolute information. Holding $\kappa$ fixed does not keep the resonant damping scale $\gamma_0$ in Eq.~\eqref{eq:g0} fixed as $\Omega$ and $\Gamma$ vary. For the baseline linewidth, moving from the baseline reservoir frequency to the lower-frequency edge of the scan increases the maximum QFI ratio but reduces the finite-time QFI evaluated at the respective ratio maxima. The largest relative gain, therefore, need not identify the setting with the largest absolute information. The scan supports the persistence and spectral tunability of the transient enhancement over the investigated settings, while showing that its magnitude and occurrence depend on the reservoir parameters.

\section{Exact solution}\label{app:exact}
\setcounter{equation}{0}

The Hamiltonian in Eqs.~\eqref{eq:H}--\eqref{eq:HI} is quadratic and conserves the total excitation number. The Heisenberg equations
\begin{align}
\dot a&=-i\omega_0a-i\lambda\sum_kg_kb_k,\nonumber\\
\dot b_k&=-i\omega_kb_k-i\lambda g_k^*a
\end{align}
are therefore linear. Their solution has the form of Eq.~\eqref{eq:aexact}, with $|G(t)|^2+\sum_k|f_k(t)|^2=1$. Eliminating the reservoir operators gives the integro-differential equation
\begin{equation}
\dot G(t)=-i\omega_0G(t)-\lambda^2\int_0^t\!\dd s\,C_0(t-s)\,G(s),\qquad G(0)=1,
\end{equation}
where $C_0(u)=\int_0^\infty\dd\omega\,J(\omega)e^{-i\omega u}$ does not depend on temperature~\cite{AnZhang2007}.

The initial reservoir state is Gaussian with zero mean, and the combination $\sum_kf_kb_k$ is a single bosonic mode in a phase-invariant Gaussian state. The reduced dynamics is therefore a thermal attenuator channel. It mixes the input with an effective thermal mode at transmissivity $\eta=|G|^2$ and adds the mean excitation $N(T,t)=\sum_k|f_k|^2n_T(\omega_k)$. The probe mean occupation is $\langle a^\dagger a\rangle(t)=\eta n_0+N$. A phase-invariant Gaussian channel is generated by a time-local master equation of the form of Eq.~\eqref{eq:me}~\cite{AnZhang2007}. Matching the evolution of the mean occupation and of the transmissivity gives the rates in Eq.~\eqref{eq:exrates}. The populations then obey Eq.~\eqref{eq:B1}, and their temperature derivatives follow from $\partial_TN$ alone, because $\eta$ is temperature independent.
We evaluate $G$ and $f_k$ numerically. We discretize the reservoir into equally spaced modes with spacing $\delta\omega$ below a cutoff $\omega_c$, with $|g_k|^2=J(\omega_k)\delta\omega$. We diagonalize the single-excitation Hamiltonian once, which gives $G(t)$ and $f_k(t)$ at all times. The discretization is accurate for times shorter than the recurrence time $2\pi/\delta\omega$.

\section{Switching envelope}\label{app:switch}
\setcounter{equation}{0}

With a time-dependent coupling $\lambda f(s)$, the integrated transition coefficients of the second-order theory become
\begin{align}
\Kt_{-}(T,t)&=\lambda^2\int_0^\infty\!\dd\omega\,J(\omega)[n_T(\omega)+1]\,|\tilde f(\omega)|^2,\nonumber\\
\Kt_{+}(T,t)&=\lambda^2\int_0^\infty\!\dd\omega\,J(\omega)n_T(\omega)\,|\tilde f(\omega)|^2,
\end{align}
with $\tilde f$ defined in Eq.~\eqref{eq:ftilde}. These expressions follow from first-order time-dependent perturbation theory for each exchange process. For $f=1$ they reduce to Eq.~\eqref{eq:Kint}, since $|\tilde f|^2=t^2\Phi_t$. The response $\chi$ follows by replacing $J(n_T+1)$ with $W_T$, and Eq.~\eqref{eq:Flead} applies unchanged. In the Markovian limit, $|\tilde f(\omega)|^2\to2\pi t_{\rm eff}\,\delta(\omega-\omega_0)$ for envelopes that vary slowly on the scale $1/\omega_0$.

For the envelope used in Sec.~\ref{sec:switch}, the rising ramp contributes
\begin{equation}
A(\Delta)=\frac12\left[E_\Delta-\frac12\left(E_{\Delta+\pi/t_r}+E_{\Delta-\pi/t_r}\right)\right],
\end{equation}
where $E_k=(e^{ikt_r}-1)/(ik)$ and $\Delta=\omega_0-\omega$. The full transform is
\begin{equation}
\tilde f(\omega)=A(\Delta)+e^{i\Delta t_r}\,\frac{e^{i\Delta(t-2t_r)}-1}{i\Delta}+e^{i\Delta t}A^*(\Delta).
\end{equation}
Each ramp contributes $3t_r/8$ to $t_{\rm eff}$, so $t_{\rm eff}=t-5t_r/4$.

\bibliography{Thernometry_References}
\end{document}